\documentclass[12pt]{article}

\usepackage{amsmath,amsfonts,amssymb}
\usepackage{graphicx}
\usepackage{psfrag}
\usepackage{enumerate}

\makeatletter \@addtoreset{equation}{section}

\makeatletter\renewcommand\section{\@startsection {section}{1}{\z@}%
                                   {-3.5ex \@plus -1ex \@minus -.2ex}%nn
                                   {2.3ex \@plus.2ex}%
                                   {\normalfont\large\bfseries}}
\renewcommand\subsection{\@startsection{subsection}{2}{\z@}%
                                     {-3.25ex\@plus -1ex \@minus -.2ex}%
                                     {1.5ex \@plus .2ex}%
                                     {\normalfont\bfseries}}

\newcommand{\be}{\begin{equation}}
\newcommand{\ee}{\end{equation}}
\newcommand{\beq}{\begin{eqnarray}}
\newcommand{\eeq}{\end{eqnarray}}

\def\[{\left [}
\def\]{\right ]}
\def\({\left (}
\def\){\right )}

\def\R{{\bf R}}

\def\CN{{\cal N}}

\def\r2{\sqrt{2}}

\def\barD{\bar{\rm D}}

\def\CH{{\cal H}}

\def\CN{{\cal N}}

\newcommand{\labell}[1]{\label{#1}\qquad_{#1}} %{\Label{#1}} %
\newcommand{\bbibitem}[1]{\bibitem{#1}\marginpar{#1}}

\def\Label#1{\label{#1}%
  \smash{\hbox to0pt{\raise1ex\hbox{\tiny[#1]}\hss}}}
\def\noLabels{\let\Label=\label}
\def\nobbibitem{\let\bbibitem=\bibitem}

\begin{document}
\noLabels % uncomment for final production
\nobbibitem % uncomment for final production

\begin{titlepage}

%\begin{flushright}%\vspace{-2cm}
%{\small
%UPR-1154-T  \\ %\vspace{-0.35cm}
%LBNL-60486 \\
%hep-th/0606118}%\\
%\end{flushright}
%\vspace{12 mm}

\vfil\
%vfil

\begin{center}

{\Large{\bf Black Ring Deconstruction\\ }}
%\vspace{3mm} {\Large{\bf From D-branes to Spacetime Foam}} \vfil

\vspace{3mm}

 Eric G.
Gimon\footnote{e-mail: eggimon@lbl.gov}$^{,a}$$^{,b}$ and Thomas S.
Levi\footnote{e-mail: levi@physics.nyu.edu}$^{,c}$
\\

\vspace{8mm}

\bigskip\medskip
\smallskip\centerline{$^a$ \it Department of Physics, University of California, Berkeley,
CA 94720, USA.}
\smallskip\centerline{$^b$
\it Theoretical Physics Group, LBNL, Berkeley, CA 94720, USA.}
\smallskip\centerline{$^c$
\it Center for Cosmology and Particle Physics}
\centerline{\it New York University, 4
  Washington Place, New York, NY 10003,  U.S.A.}

\vfil

\end{center}
\setcounter{footnote}{0}
%%%%%%%%%%%%%%%%%%%%%%%%%%%%%%%%%%%%%%%%%%%%%%%%%%%%%%%%%%%%%%%%%%%%%%%%%%%%%%%%%%%%%%%
\begin{abstract}
\noindent
We present a sample microstate for a black ring in four and five dimensional
language. The microstate consists of a black string microstate with an
additional D6-brane. We show that with an appropriate choice of parameters the
piece involving the black string microstate falls down a long $AdS$ throat,
whose M-theory lift is $AdS_3 \times S^2$. We wrap a spinning dipole M2-brane
on the $S^2$ in the probe approximation. In IIA, this corresponds to a
dielectric D2-brane carrying only D0-charge. We conjecture this is the first
approximation to a cloud of D0-branes blowing up due to their non-abelian
degrees of freedom and the Myers effect.
\end{abstract}
%%%%%%%%%%%%%%%%%%%%%%%%%%%%%%%%%%%%%%%%%%%%%%%%%%%%%%%%%%%%%%%%%%%%%%%%%%%%%%%%%%%%%%%%%
\vspace{0.5in}

\end{titlepage}
\renewcommand{\baselinestretch}{1.05}  %Line spacing
%%%%%%%%%%%%%%%%%%%%%%%%%%%%%%%%%%%%%%%%%%%%%%%%%%%%%%%%%%%%%%%%%%%%%%%%%%%%%%%%
%%%%%%%%%%%%%%%%%%%%%%%%%%%%%%%%%%%%%%%%%%%%%%%%%%%%%%%%%%%%%%%%%%%%%%%%%%%%%%%%%%%%%%%%%%%
%\tableofcontents

\newpage

\section{Introduction}

One of the great successes of string theory has been the explanation of black
hole entropy in terms of underlying microstates of string solitons
\cite{sen,stromingervafa}. These analyses apply at weak coupling when there is
no macroscopic horizon. This picture is somewhat lacking, in that we have no
understanding of how a macroscopic horizon can emerge. Recently, a second
approach has appeared (see \cite{mathurreview,mathur2} for reviews) where for
sufficiently supersymmetric black holes at least some of the microstates can be
viewed as a complicated ``spacetime foam''
\cite{mathurreview,mathur2,us,bw2,4dmicro}. In this picture, the black hole
with a macroscopic horizon is the effective semiclassical description of the
foam. Using ideas from \cite{denef1,denef2,denef3,denefhall} it was shown in
\cite{4dmicro} how to connect these two pictures. As the string coupling grows,
D-brane bound states grow a transverse size, leading to a topological
description as a spacetime foam. A further conjecture was made in
\cite{4dmicro}, namely that every supersymmetric four-dimensional black hole of
finite area, perserving 4 supercharges, can be split up into microstates of
primitive 1/2-BPS ``atoms'', each of which preserves 16 supercharges. To
describe a bound state, these atoms should have mutually non-local charges.

The 1/2-BPS atoms can be viewed as type IIA branes (generally with worldvolume
fluxes turned on) wrapping the internal manifold. A ``scaling'' solution is
present when the charges are chosen such that the corresponding quiver quantum
mechanics of open strings on these branes has a closed loop. In \cite{bwmerger}
it was shown that this scaling causes the branes making up the closed loop to
fall down a long AdS throat of constant cross-sectional area. In \cite{4dmicro}
it was further claimed that the non-Abelian degrees of freedom of each stack of
branes will be important in giving the black hole its finite entropy and size.
A significant step was taken in this direction in \cite{andydecon}, where it
was shown that one could add D0-brane charge in the probe approximation by
wrapping a dipole D2-brane around a certain $S^2$ in the non-compact space. It
was conjectured that this could be due to the Myers effect causing a cloud of
D0-branes to puff up to the dipole D2-brane.

When we lift the four-dimensional IIA solutions to M-theory in five dimensions,
there is a greater variety of black objects available to us: strings, holes and
rings \cite{reall1}. The example microstate constructed in \cite{andydecon} was
that of a black string in five dimensions. In this short note, we will show
that a similar construction can be carried out for a black ring microstate. In
four dimensions this becomes the black string microstate with an additional
D6-brane (just as a black ring in Taub-Nut becomes a black hole with an
additional D6-brane \cite{EEMR}). We find we can indeed add D0-brane charge by
wrapping a dipole D2-brane. The quantization of these configurations may
dominate the entropy for the corresponding black ring.

In section 2 we review the construction of microstates and the connection
between four and five dimensions. In section 3 we give our sample black ring
microstate and find a scaling solution. We show that the closed loop part of
the configuration falls down a long AdS throat, and the remaining D6-brane
remains outside this throat. In section 4 we find a hierarchy of scales inside
the throat region and that the metric simplifies in both a ``near'' and ``far''
region. In section 5, we show we can wrap a dipole membrane and generate
angular momentum in five dimensions which becomes D0-brane charge in four
dimensions. We conclude with some discussions and future directions in section
6.

\section{Basics and setup}

In this section, we will briefly review our framework for finding microstate
solutions.  In \cite{us,bw2}, a general framework was developed for microstate
solutions in five non-compact dimensions. In \cite{4dmicro} these solutions
were extended to four dimensions, with IIA compactified on $T^6$ give solutions
to $\CN=8$ supergravity. In the STU sector, with branes charges dual to the
heterotic charges in \cite{STU}, the general solution can be completely
characterized by a symplectic vector of eight harmonic
functions on $\R^3$%
\beq%
{\cal H} = (M_0, M_i, K^0, K^i) = \Gamma_\infty+\sum_p {\Gamma_p \over
\rho_p}, \quad i=1,\ldots,3,
\eeq%
where $\rho_p = | \vec{x} - \vec{x}_p|$ for some positions
$\vec{x}_p$.  $\Gamma_\infty=(\delta M_0, \delta M_i, \delta K^0, \delta K^i)$
is a constant vector, which completely determines the values of the
scalars at infinity. It must satisfy:
\be
\label{BC}
J_4(\Gamma_\infty) = 1, \qquad <\Gamma_\infty,\sum_p \Gamma_p> = 0
\ee
where $J_4$ is the quartic invariant of $E_{7(7)}$ applied to our restricted
set of charges.  For a pair of general vectors of the form
$\Gamma_p=(Q_0^p,Q_i^p,Q^0_p,Q^i_p)$ the U-duality invariant symplectic product
is defined as (sum on $i$ implied)
\be%
<\Gamma_p,\Gamma_q> \, =\, {1\over 2} \(Q^0_p Q_0^q - Q^0_q Q_0^p + Q^i_p Q_i^q
- Q^i_q Q_i^p\) .
 \Label{integerint}
\ee%
 Each center represents a D-brane wrapping a cycle in the internal
manifold with worldvolume fluxes turned on and quantized charges $(D6, D2_i,
D0, D4^i)$ given by $\Gamma_p$. Throughout this paper we will take the D2 and
D4 charges to be diagonal in the $i$ index, and merely identify them with a
$1$, e.g. $M_i = M_1, \ \forall i$.

Using the harmonic function, we can construct several
combinations which will appear in our expressions throughout. We define%
\beq%
Z &=& M_1-{(K^1)^2 \over M_0} ,\\
k_0 &=& {L\over 4} K^0 +{L\over 2} {(K^1)^3 \over M_0^2}-{3L\over 4} {M_1 K^1 \over M_0}, \\
J_4 &=& {L^2 \over 4} \left(Z^3 H-k_0^2 H^2 \right), \quad H=-{4\over L^2} M_0 ,
\eeq
where $L$ is the radius of the M-theory circle (see \cite{4dmicro}), which we
will set equal to one from now on along with $8G_N^{(4)} = 1$.

The ten-dimensional (four non-compact) IIA string frame metric and dilaton are given by
\footnote{There are non-trivial gauge and B-fields in this background as well, we omit them
here for brevity's sake. For details we refer the reader to \cite{4dmicro}.}%
\beq%
ds^2_{10} &=& -J_4^{-1/2} (dt +k_a dx^a)^2+J_4^{1/2}
\left(ds^2_{\R^3}+ (-Z M_0)^{-1} ds^2_{T^6}\right) , \\
e^{2\Phi} &=& (J_4)^{3/2} (-Z^3 M_0 ^3)^{-1} .
\eeq%
where $k_a dx^a$ is a one-form that satisfies%
\be%
\star_3 d(k_a dx^a) =<d \CH,\CH> ,%
\ee%
and the Hodge star operates on the flat $\R^3$ only. There is an integrability condition
on $k_a dx^a$, which leads to the bubble equations \cite{4dmicro,us,bw2}%
\beq%
<\Gamma_p, \Gamma_\infty> +\sum_{q\neq p} {<\Gamma_p,\Gamma_q> \over \rho_{pq}} =0 ,%
\eeq%
where $\rho_{pq} =|\vec{x}_p - \vec{x}_q|$. The sum of these equations just
reduce to the condition in eq.(\ref{BC}).  Note also that the asymptotic volume
of $T^6$ is proportional to $(-ZM_0)^3$.  There are two possible conventions
here: one can either normalize $T^6$ to unit volume and let the asymptotic
value of $(-ZM_0)$ determine its volume, or one can restrict the choice of
$\Gamma_\infty$ further such that this expression asymptotes to one and let
$T^6$ have arbitrary volume.  We choose the latter convention, with the
additional convenient choice of all equal size $T^2$'s.

When we lift our ten-dimensional metric to M-theory we obtain%
\beq%
ds^2_{11} &=& -Z^{-2}(dt+k_0 \sigma+k_a dx^a)^2 +Z \,ds^2_{HK}+ds^2_{T^6}, \\%
ds^2_{HK} &=& H^{-1} \sigma^2 + H (dr^2 +r^2 d\theta^2+r^2 \sin^2\theta
d\phi^2) = H^{-1} \sigma^2 + H d\tilde{s}^3_{\R^3} ,
\eeq%
where $r = {L\,\rho\over 2} = \rho/2$. The one-form $\sigma=d\tau +f_a dx^a$ is
defined by $d \sigma = \star_3 dH$.  Note that if $\lim_{r\to\infty} k_0 =
{1\over 2}\tan\alpha\sec\alpha \ne 0$, i.e. if the total central charge of our
solution has a non-zero phase $\alpha$, our lift puts the M-theory metric in
the somewhat unorthodox form where $\partial_t$ and $\partial_\tau$ are no
longer orthogonal at infinity (see \cite{EEMR} for a similar discussion).  For
$\alpha \to \pm\pi/2$, the asymptotic metric goes to $\mp dt\,\sigma + {1\over
4} \sigma^2 + 4r^2ds^2_{\mathbb S^2}$; the vector $\partial_t$ becomes null as
it does for all the other zeros of $H$ at finite $r$.

\section{A sample black ring microstate}
In this section, we will construct an example that represents the semiclassical
limit of a black ring microstate. In \cite{andydecon} a sample microstate for a
D4-D0 black hole was found. When lifted to five dimensions, this solution
becomes a wrapped black string. To get to a black ring microstate we add a
single (pure) D6-brane that is separated by a parametrically large distance
from the initial branes in \cite{andydecon}. The addition of the D6-brane means
that the M-theory circle that the would-be black string (whose microstate this
is) wraps is now contractible, turning this solution into a black ring
microstate where the radius of the ring is proportional to the distance to the
added D6-brane. If this distance were to become parametrically small, the extra
pole becomes part of our ``fuzzball"  and our solution would be a black hole
microstate. All solutions lift to mostly smooth solutions in eleven dimensions
characterized by a ``foam'' of two-cycles \cite{us,bw2} with a shock-wave from
the D0-branes.

We begin by placing the initial branes. The D6 and
$\barD 6$-branes  carry worldvolume flux which induces the other brane charges.
They can be written%
\beq%
\Gamma_1={1\over2}(-1, -m^2, m^3, m), \quad \Gamma_2={1\over2}(1,m^2,m^3,m) \, .%
\eeq%
We also add $q$ D0-branes with total charge vector%
\be%
\Gamma_3 = {q\over2}(0,0,-1,0) .%
\ee%
Lastly, we add a single, pure D6-brane with charge vector%
\be%
\Gamma_0 = {1\over2}(-1,0,0,0) .%
\ee%
In \cite{us} it was shown how to construct a quiver model for the corresponding
open string quantum mechanics living on these branes, Fig. 1 displays the
quiver for this setup, the numbers on the diagonals are the intersection
numbers $\gamma_{pq} = <\!\Gamma_p\,, \Gamma_q\!>/G_N^{(4)}$.  Note that each
node is associated with a $U(1)$ gauge group, except the third note which has
gauge group $U(q)$ and thus non-Abelian degrees of freedom.

\begin{figure}
\centering \hspace{0.2in}
\includegraphics[width=0.4\textwidth]{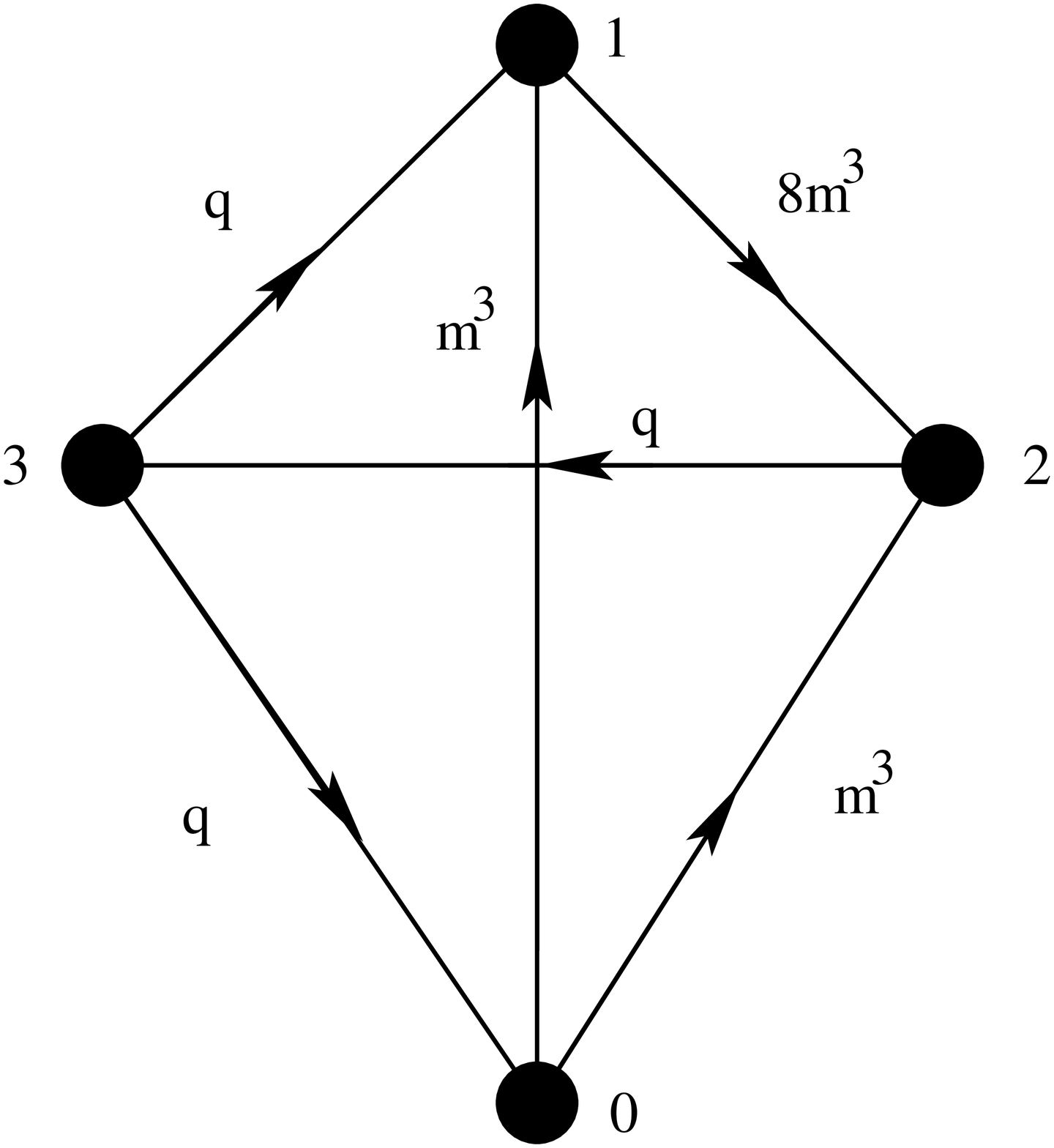} \caption{The quiver for
the black ring microstate example.}
\end{figure}

It was also shown in \cite{4dmicro,denefhall} that whenever a given quiver has
a closed loop, so long as the corresponding intersection numbers satisfy
triangle inequalities, we can find a ``scaling'' solution for that part of the
quiver. These scaling solutions allow the distance between the corresponding
branes {\it in the flat $\R^3$} base to be taken arbitrarily small. In the full
metric, instead we develop a longer and longer throat region which is smoothly
capped. The cross-sectional area of this cap remains constant as the throat
deepens \cite{bwmerger}. Looking at the quiver, we can see there are two
possible closed loops, 1-2-3 and 2-3-0. We will analyze the 1-2-3 closed loop
here, the 2-3-0 case can be solved in a similar manner.

Grouping terms suggestively, the bubble equations become (letting the positions
of the individual D0-branes be labelled $\vec{x}_{3_i}$)%
\beq%
\left( <\Gamma_1, \Gamma_\infty> - {m^3 \over 8 \rho_{10} } \right)
&+& \left( {m^3\over\rho_{12} }- \sum_{i=1}^q{1 \over 8 \rho_{13_i} }\right) =0 , \labell{bubble1}\\
\left( <\Gamma_2, \Gamma_\infty> - {m^3 \over 8 \rho_{20} } \right)
&+& \left( -{m^3\over\rho_{12} } + \sum_{i=1}^q {1 \over 8 \rho_{23_i} }\right) =0 , \labell{bubble2} \\
\left( <{\Gamma_3\over q}, \Gamma_\infty> + {1 \over 8 \rho_{03_i} } \right)
&+& \left( {1 \over 8 \rho_{13_i} }- {1 \over 8 \rho_{23_i} }\right) =0 , \quad \forall i=1,\ldots,q. \labell{bubble3} %
\eeq%
We are seeking a solution where $\rho_{10},\rho_{20},\rho_{03_i} \to \rho_{0}
\gg \rho_{12}, \rho_{13_i}, \rho_{23_i} \to 0$. That is, in each of the bubble
equations, the terms in the second set of parentheses will dominate while the
first term drops out just as the constant terms were ignored in such equations
for the black string microstate in \cite{andydecon}. The last equation
\eqref{bubble3} tells us that each D0-brane must be equidistant between the D6
and $\barD 6$-brane, $\rho_{13_i}=\rho_{23_i}$, so the D0-branes lie on a plane
between them.
Using this in \eqref{bubble1} we find that the separation of the D6 and $\barD 6$ pair is%
\be%
\rho_{12} = 8m^3 \left( \sum_{i=1}^q {1\over \rho_{13_i}}\right)^{-1} \equiv 2 R_6 .%
\ee%
We still need to make sure the triangle inequalities are satisfied for this to be a solution.
The distance from any D0-brane to either D6-brane in the pair is at least $R_6$,
so to satisfy the triangle inequalities we require%
\be%
q \geq 4m^3.%
\ee%
When this is satisfied we have a scaling solution.  It is interesting to note
that this is different from, but implies, the condition for $J_4(\Gamma_1 +
\Gamma_2 + \Gamma_3) \geq 0$ which is $q\geq 2m^3$.  In this case, the branes
making up the closed loop part of the quiver fall down a long $AdS$ throat,
with the remaining D6-brane left outside. Fig. 2 shows a rough sketch of this.

\begin{figure}
\centering \hspace{0.2in}
\includegraphics[width=0.6\textwidth]{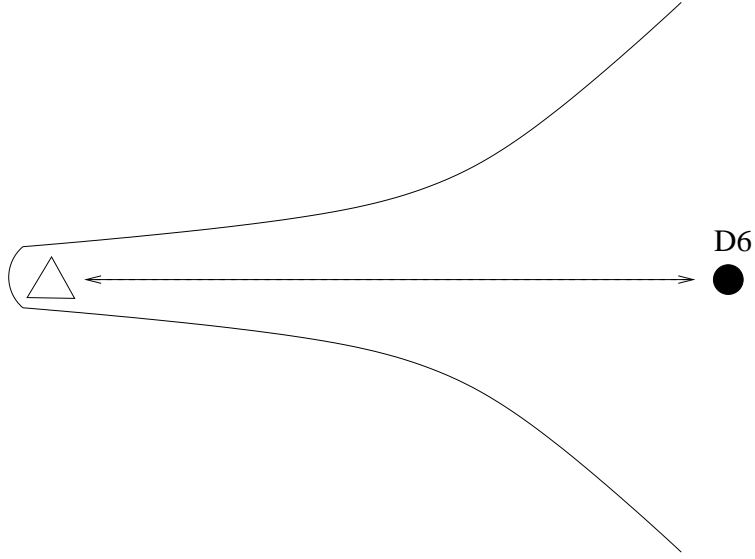} \caption{Rough sketch
of the closed loop of the quiver falling down a throat in the scaling solution
with a single D6-brane left outside.}
\end{figure}

We can determine an effective $\R^3$ distance from the pure D6-brane to the
other three D-branes (the real geodesic distance scales to infinity as we bring
those last three poles together) by aggregating these three in $\Gamma_R =
\Gamma_1 + \Gamma_2 + \Gamma_3$ and using a simple two-pole constraint equation
(we restore $G^{(4)}_N,L$ here):
\be%
\rho_{0} \approx  - {<\Gamma_0,\Gamma_R>\over <\Gamma_0, \Gamma_\infty>} =
{4G^{(4)}_N\, (q - 2m^3)\over L\, \delta K^0} .
\Label{ringradius}%
\ee%
This number only makes physical sense if the asymptotic moduli are such that
$\delta K^0 > 0$, otherwise the SUSY solution disappears . In the scaling
limit, $\rho_0$ is by definition much larger than the decoupled scales
$\rho_{12},\rho_{13} $ and $\rho_{23}$.  Thus we have the beginnings of a nice
hierarchy of scales.

\section{Extending the hierarchy of scales}
We have seen that we can find a scaling solution for a black ring microstate,
in which the D6-$\barD 6$-D0 branes fall down a long throat, and a single
D6-brane remains outside. If we focus on the region down the throat (the closed
loop part of the quiver), then this case becomes the black string case of
\cite{andydecon}. We can take $q\gg 4m^3$, in which case the D6-branes will
typically be much closer to the plane containing the D0-branes, than to any of
the D0-branes themselves. As a simple example, we can let all of the D0-branes
lie in a ring of radius $R_0$. We then find
\be%
R_6 = {4m^3 \over q} \sqrt{R_6^2+R_0^2} \ll R_0 .%
\ee%
As in \cite{andydecon} we find that the metric simplifies in both a far region,
$\rho_0 \gg |\vec{x}| \gg R_0$ and a near region $|\vec{x}| \ll R_0$. In the far region,
the metric simply becomes that of the near horizon of a D4-D0 black hole, whose
lift is
\be%
ds^2_5 \approx {(q-2m^3)\over 8m} d\tau^2 -{2r \over m} d\tau dt+m^2 \left(
{dr^2\over r^2}+d\eta^2+\sin^2\eta d\psi^2 \right) .
\ee%
In the near region, we can change to prolate spheroidal coordinates%
\be%
\rho_1 =R_6 (\cosh\beta+\cos\eta), \quad \rho_2=R_6 (\cosh\beta -\cos\eta) ,%
\ee%
and further make the coordinate transformations%
\be%
t={2m^3 \over R_6 }x, \quad \tau=2(x+\theta), \quad \phi=\psi+x-\theta ,%
\ee%
to obtain%
\be%
ds_5^2 \approx m^2(d\eta^2+\sin^2\eta d\psi^2)+4m^2
\left(-\cosh^2({\beta\over2})dx^2+\sinh^2({\beta\over2})d\theta^2+{1\over4} d\beta^2\right),
\Label{adsmetric}%
\ee%
which is global $AdS_3 \times S^2$. Note that we have regular $2\pi$
identifications on $\theta$ and $\psi$.

Let us step back a moment and review what we have done. We started with a
solution for a black ring microstate that consisted of two pieces, a cloud of
D6-$\barD 6$-D0 branes and a single pure D6-brane separated from the cloud. We
found that there was a scaling solution, where the cloud of branes fell down a
long throat, leaving the single D6-brane outside. We then focused in on the
cloud of branes down the throat and found this case to be that of the black
string microstate analyzed in \cite{andydecon}. In particular, if we take $q\gg
4m^3$, we find that the ring of D0-branes is much further from the pair of
D6-branes than the D6-branes are from themselves, which allows us to focus in
further on both a near and far limit. Hence, we should be able to
``deconstruct'' the black ring in a similar way to that of the black string.

\section{Wrapped branes and deconstructing the ring}

In this section, we show that in the probe approximation, we can add a dipole
M2-brane that wraps the $S^2$ that appears down the throat and in the near
limit. Because the cycle it wraps is in a contractible class at infinity, it cannot
carry any net M2 charge. This dipole brane does carry angular momentum around a circular
direction in the $AdS_3$, which when reduced along $\tau$ leads to a D0-brane
charge. We conjecture that this is a result of the Myers effect on a cloud of
D0-branes when brought near the pair of D6-branes, where gradients of both the
dilaton and RR one-form play a role.

In \cite{andydecon} it was shown for the black string case that in the near
limit one can wrap a supersymmetric M2-brane on the $S^2$ of \eqref{adsmetric},
sitting at constant $\beta=\beta_0$ and $\tau=\tau_0$ (and hence moving in
$\theta=\tau_0/2 - x$). This is an ellipsoidal brane with the D6-$\barD 6$ pair
acting as the foci for the ellipsoid. We will wrap our M2-brane in the same way
for the black ring case. It was further claimed that this configuration would
be BPS in the full geometry since in the black string case in both the near
limit and the full geometry the base space depends only on $M_0$, which is the
same for both (for convenience the constant term in $M_0$ was set to zero, i.e
$\alpha$ was set to $\pm\pi/2$). Here, there is a further subtlety, since in
the full geometry, $M_0$ includes the added pure D6-brane. However, we do not
anticipate a large change because the probe brane is being added very far down
the throat and in the probe approximation.

One can readily calculate the angular momentum around the $\theta$
circle (we use $\tau_{M2} = 1/(4 \pi^2 l_p^3)$ and return other dimensionful
factors) contributed by the M2-brane
\be
J_\theta = -{2 m^3 \over \pi} \,{(2\pi l_p)^6\over V_{T^6}} \sinh^2 ( {\beta_0 \over 2} ) .
\ee
While this will correspond to adding positive D0-brane charge upon reduction,
it reduces the net $J_R$ of our solution.  This reduction typically increases
the entropy of SUSY black objects in five dimensions, so adding wrapped objects
like this M2 seems like an efficient way to provide new solutions for
microstates of finite entropy black objects in five dimensions.

We wish to reduce this setup to get a IIA solution. One subtlety that arises is
precisely which circle to reduce on, along $\partial_\theta$ or
$2\partial_\tau$?  While the Killing vector $\partial_\theta$ appears to be the
natural choice in the near metric, in the full geometry the reduction is along
$2\partial_\tau$, and it is thus this circle that we should use to
appropriately define our charges. One might well ask how a reduction along
$2\partial_\tau$ produces any D0-brane charge in four dimensions, since the
M2-brane position is independent of $\tau$. The answer is that the
$\tau$-circle both varies in size and is non-trivially fibered over the $S^2$
that the M2-brane wraps; these two features each contribute to the $U(1)$ field
strength $F$ on the D2-brane (see \cite{bergtown} for more details on the
reduction). Dualizing the appropriate vielbein in the usual way gives us the
worldvolume field on the D2-brane
\be%
{\cal F} = F =  {m^3\over \pi} \sin \eta \left( \sinh^2 ({\beta_0 \over 2})
d\eta \wedge d\varphi-d\eta \wedge dx \right)\({(2\pi l_p)^6\over V_{T^6}}\) ,%
\ee%
where $\varphi={\tau \over 2}-x$. This dipole D2-brane carries only net
quantized D0-charge
\be%
n_{D0} = - J_\theta, %
\ee%
and no other charges, in particular one can see that the fundamental string
charge vanishes because of the simple topological fact that the M2-brane is not
wrapping the $\tau$ direction. To get the physical D0-charge (i.e. to establish
the D0-brane tension) we need a value for $g_s$, set by the radius of
compactification. In the region near our closed quiver, in both the near and
far limit, we are operating in a decoupling regime which has ``forgotten'' any
scale relating to the size of the $\tau$ circle: there is no set value we can
use.  In the full geometry, one could choose as a reference scale the
separation from the naked D6 brane in (\ref{ringradius}) or the asymptotic size
of the M-circle $L$.

Finally, as in \cite{andydecon}, these branes move on the
$T^6$ as particles in a magnetic field due to the C and B-fields with
legs on the torus. It seems possible to add D0-brane charge to a
microstate by using these wrapped branes, which could be interpreted
as a cloud of D0-branes blown up by the Myers effect. However, we
caution that this calculation was done in the probe approximation. It
would be interesting to attempt to take into account the backreaction.

\section{Discussion}
In this note we have presented an example of a black ring microstate and shown
that a picture emerges of a cloud of branes falling down a long AdS throat, and
a single brane remaining outside. In the throat region one can take a near and
far limit. In the near limit the D0-branes are distant actors whose only role
is to set the overall AdS scale. We then showed that we can add some more
D0-brane charge back by wrapping a dipole D2-brane and conjectured that the
complete cloud of D0-branes would blow-up into such a dipole brane via the
Myers effect, much like giant gravitons. It would be interesting to explore
detailed aspects of this phenomenon.

  To nail down the physics of microstates in detail we will need to go
beyond the assumptions in this paper.  Our construction of the wrapped brane
was done in the probe limit, and only in the near region. In particular, it is
not clear to us why a dipole brane carrying all the D0 charge would have such a
uniform charge density on $S^2$ so unlike the original charge density of the
D0-brane ring confined to the equatorial plane of $S^2$. It would be useful to
see how the dielectric construction carries over when the backreaction of the
charged D2-brane is taken into account. In the M-theory picture perhaps more
general pseudo-Hyper-Kahler spaces will appear (see \cite{BWnew} for some
examples of such).  Additionally, can we actually see a cloud of D0-branes
puffing up by using their non-Abelian degrees of freedom?  One might have
expected a ring of D0-branes to blow up into more of a torus than a sphere.  In
general, we feel that it will be essential to understand how these di-electric
effects act on component branes to fully understand how much of the entropy a
black object is accounted for by the scaling solutions containing a small
number of nodes. Do these small quivers dominate the ensemble for the
corresponding black object?

Finally, the closed loop in our example and that of \cite{andydecon} has
several particular properties which may not persist for more general closed
triples of nodes.  First, one of the nodes, the D0-branes, has zero
intersection number with the sum over the other two.  It would be interesting
to understand what happens if were to add a little extra D2-charge to this
node: a dielectric D4-brane perhaps?  Second, only two of the nodes in the
closed loop are primitive, they cannot be split into smaller integer quantized
stacks and so have no non-Abelian degrees of freedom.  With all three nodes
non-primitive, could we have multiple and simultaneous dielectric effects?
Again, the question arises as to how generic our example is.  It certainly
cannot be a microstate for its "parent" black object if $2m^3 > Q^0 = (q -
2m^3)> 0 $. To get a full picture of what are typical microstates and what
dominates the entropy of black objects, we either need to extend our basic
picture to a more general class of quivers, or understand why these other
candidates don't contribute in a substantial way to the entropy of the
ensemble.

These and other issues present challenging problems, but solving them
is important for our understanding of black objects, the information
puzzle and how classical spacetimes emerge from underlying quantum
mechanical states.

\section*{Acknowledgements}
We thank Vishnu Jejjala, Matt Kleban and Joan Simon for
discussions. We also thank Vijay Balasubramanian for collaboration
during the early stages of this project. We are grateful to the
organizers of the The Sowers Theoretical Physics Workshop 2007 at
Virginia Tech, where part of this work was conducted. TSL thanks UC
Berkeley for hospitality during some of this work.  The work of EGG was
supported by the US Department of Energy under contracts DEAC03-
76SF00098 and DE-FG03-91ER-40676, the National Science Foundation
under grant PHY-00-98840 and the Berkeley Center for Theoretical Physics.

\bibliographystyle{utphys}

\bibliography{DW}

\providecommand{\href}[2]{#2}\begingroup\raggedright\begin{thebibliography}{10}

\bibitem{sen}
A.~Sen, ``Extremal black holes and elementary string states,'' {\em Mod. Phys.
  Lett.} {\bf A10} (1995) 2081--2094,
\href{http://arXiv.org/abs/hep-th/9504147}{{\tt hep-th/9504147}}.
%%CITATION = HEP-TH 9504147;%%.

\bibitem{stromingervafa}
A.~Strominger and C.~Vafa, ``Microscopic origin of the bekenstein-hawking
  entropy,'' {\em Phys. Lett.} {\bf B379} (1996) 99--104,
\href{http://arXiv.org/abs/hep-th/9601029}{{\tt hep-th/9601029}}.
%%CITATION = HEP-TH 9601029;%%.

\bibitem{mathurreview}
S.~D. Mathur, ``The fuzzball proposal for black holes: An elementary review,''
\href{http://arXiv.org/abs/hep-th/0502050}{{\tt hep-th/0502050}}.
%%CITATION = HEP-TH 0502050;%%.

\bibitem{mathur2}
S.~D. Mathur, ``The quantum structure of black holes,''
\href{http://arXiv.org/abs/hep-th/0510180}{{\tt hep-th/0510180}}.
%%CITATION = HEP-TH 0510180;%%.

\bibitem{us}
P.~Berglund, E.~G. Gimon, and T.~S. Levi, ``Supergravity microstates for bps
  black holes and black rings,''
\href{http://arXiv.org/abs/hep-th/0505167}{{\tt hep-th/0505167}}.
%%CITATION = HEP-TH 0505167;%%.

\bibitem{bw2}
I.~Bena and N.~P. Warner, ``Bubbling supertubes and foaming black holes,''
\href{http://arXiv.org/abs/hep-th/0505166}{{\tt hep-th/0505166}}.
%%CITATION = HEP-TH 0505166;%%.

\bibitem{4dmicro}
V.~Balasubramanian, E.~G. Gimon, and T.~S. Levi, ``Four dimensional black hole
  microstates: From d-branes to spacetime foam,''
\href{http://arXiv.org/abs/hep-th/0606118}{{\tt hep-th/0606118}}.
%%CITATION = HEP-TH 0606118;%%.

\bibitem{denef1}
F.~Denef, ``Supergravity flows and d-brane stability,'' {\em JHEP} {\bf 08}
  (2000) 050,
\href{http://arXiv.org/abs/hep-th/0005049}{{\tt hep-th/0005049}}.
%%CITATION = HEP-TH 0005049;%%.

\bibitem{denef2}
F.~Denef, B.~R. Greene, and M.~Raugas, ``Split attractor flows and the spectrum
  of bps d-branes on the quintic,'' {\em JHEP} {\bf 05} (2001) 012,
\href{http://arXiv.org/abs/hep-th/0101135}{{\tt hep-th/0101135}}.
%%CITATION = HEP-TH 0101135;%%.

\bibitem{denef3}
B.~Bates and F.~Denef, ``Exact solutions for supersymmetric stationary black
  hole composites,''
\href{http://arXiv.org/abs/hep-th/0304094}{{\tt hep-th/0304094}}.
%%CITATION = HEP-TH 0304094;%%.

\bibitem{denefhall}
F.~Denef, ``Quantum quivers and hall/hole halos,'' {\em JHEP} {\bf 10} (2002)
  023,
\href{http://arXiv.org/abs/hep-th/0206072}{{\tt hep-th/0206072}}.
%%CITATION = HEP-TH 0206072;%%.

\bibitem{bwmerger}
I.~Bena, C.-W. Wang, and N.~P. Warner, ``Mergers and typical black hole
  microstates,'' {\em JHEP} {\bf 11} (2006) 042,
\href{http://arXiv.org/abs/hep-th/0608217}{{\tt hep-th/0608217}}.
%%CITATION = HEP-TH 0608217;%%.

\bibitem{andydecon}
F.~Denef, D.~Gaiotto, A.~Strominger, D.~Van~den Bleeken, and X.~Yin, ``Black
  hole deconstruction,''
\href{http://arXiv.org/abs/hep-th/0703252}{{\tt hep-th/0703252}}.
%%CITATION = HEP-TH/0703252;%%.

\bibitem{reall1}
R.~Emparan and H.~S. Reall, ``A rotating black ring in five dimensions,'' {\em
  Phys. Rev. Lett.} {\bf 88} (2002) 101101,
\href{http://arXiv.org/abs/hep-th/0110260}{{\tt hep-th/0110260}}.
%%CITATION = HEP-TH 0110260;%%.

\bibitem{EEMR}
H.~Elvang, R.~Emparan, D.~Mateos, and H.~S. Reall, ``Supersymmetric 4d rotating
  black holes from 5d black rings,''
\href{http://arXiv.org/abs/hep-th/0504125}{{\tt hep-th/0504125}}.
%%CITATION = HEP-TH 0504125;%%.

\bibitem{STU}
M.~J. Duff, J.~T. Liu, and J.~Rahmfeld, ``Four-dimensional string-string-string
  triality,'' {\em Nucl. Phys.} {\bf B459} (1996) 125--159,
\href{http://arXiv.org/abs/hep-th/9508094}{{\tt hep-th/9508094}}.
%%CITATION = HEP-TH/9508094;%%.

\bibitem{bergtown}
E.~Bergshoeff and P.~K. Townsend, ``Super d-branes,'' {\em Nucl. Phys.} {\bf
  B490} (1997) 145--162,
\href{http://arXiv.org/abs/hep-th/9611173}{{\tt hep-th/9611173}}.
%%CITATION = HEP-TH/9611173;%%.

\bibitem{BWnew}
I.~Bena, N.~Bobev, and N.~P. Warner, ``Bubbles on manifolds with a u(1)
  isometry,''
\href{http://arXiv.org/abs/arXiv:0705.3641 [hep-th]}{{\tt arXiv:0705.3641
  [hep-th]}}.
%%CITATION = ARXIV:0705.3641;%%.

\end{thebibliography}\endgroup

\end{document}